\documentclass[pre,preprint,groupedaddress,a4paper,showpacs]{revtex4}

\usepackage{natbib}
\usepackage[Large]{subfigure}
\usepackage{dcolumn}
\usepackage[dvips]{graphicx}
\usepackage{amsmath}
\usepackage{bm}
\usepackage{units}

\begin{document}

\newcommand{\degrc}{\ensuremath{^{\circ}}C}

\bibliographystyle{apsrev}

\title[Microrheology]{Colloidal dynamics in polymer solutions: Optical two-point microrheology measurements}

\author{Laura Starrs}
\author{Paul Bartlett}
\email{P.Bartlett@bristol.ac.uk}
 \affiliation{School of Chemistry, University of Bristol, Bristol
BS8 1TS, UK.}

\date{\today}

\begin{abstract}

We present an extension of the two-point optical
microrheology technique introduced by Crocker \textit{et
al.} [Phys. Rev. Lett. \textbf{85}, 888 (2000)] to high
frequencies. The correlated fluctuations of two probe
spheres held  by a pair of optical tweezers within a
viscoelastic medium  are determined using optical
interferometry. A theoretical model is developed to yield
the frequency-dependent one- and two-particle response
functions from the correlated motion. We demonstrate the
validity of this method by determining the one- and
two-particle correlations in a semi-dilute solution of
polystyrene in decalin. We find that the ratio of the one-
and two-particle response functions is anomalous which we
interpret as evidence for a slip boundary condition caused
by depletion of polymer from the surface of the particle.

\end{abstract}
\pacs{82.70.Dd, 83.50.Fc, 05.40.+j}

\maketitle

\section{Introduction}
\label{sec:intro}

The dynamics of colloidal particles dispersed in
viscoelastic polymer solutions are important in both
fundamental and applied science. Suspensions of particles
in polymers are encountered in an vast range of chemical
products including coatings, controlled-release drug
formulations, personal care products, and filled polymer
composites as well as being significant in many biological
processes, such as intracellular transport inside the
viscoelastic cytoplasm of a cell \cite{1837}. From a
fundamental viewpoint, the characteristic lengths of
polymers and colloids which differ by several orders of
magnitude, suggest that these mixtures should be ideal
candidates for coarse-grained descriptions. However while
such approaches have been applied  to equilibrium
properties with considerable success there have been no
comparable attempts, so far, to explain and quantify the
rich dynamical and rheological properties of these complex
mixtures. Understanding the dynamics of these complex
systems remains an important goal of soft matter science.
In recent years considerable experimental progress in the
characterization of these systems has been made though the
development of optical microrheology techniques. The underlying
idea, pioneered by Mason and Weitz \cite{655}, is that the
Brownian fluctuations of a colloidal probe particle reflect the
viscoelasticity of the medium in which the probe is embedded. By
analysing the thermal motion it is possible to obtain quantitative
information about the rheological properties of the polymer matrix
over an extended range of frequencies not accessible to
conventional rheometers. Optical microrheology experiments
typically monitor the thermal fluctuations of a \textit{single}
probe particle using a combination of video microscopy, optical
interferometry \cite{1169,1170}, or diffusing wave spectroscopy
\cite{655}. While microrheological techniques enjoy significant
potential advantages over traditional rheological methods
complications in interpretation have limited their application
\cite{2302}. The problem is that the presence of a particle can
perturb the surrounding medium (by, for instance, depleting
polymer segments from near the particle surface) so that the
dynamics of a single particle fails to reflect the macroscopic
viscoelasticity of the medium \cite{1545,1572}. In effect, a
particle diffuses in a ``pocket'' of material whose local
rheological properties are not those of the bulk. To circumvent
this limitation, Crocker \textit{et al.} \cite{1572} conjectured
that the correlated fluctuations of \textit{two} separated probe
particles should measure the bulk rheological properties of
inhomogeneous materials  more accurately than traditional
one-particle experiments. Using video microscopy they tracked the
thermal diffusion of pairs of probe particles embedded in a
viscoelastic solution of guar and confirmed that the extracted
viscoelastic moduli were in close agreement with conventional
measurements. Significantly, the single particle results differed
substantially, both in magnitude and scaling, from the
two-particle results. Recently these observations have been
confirmed theoretically by Levine and Lubensky \cite{1547}. Using
an analogy to classical electrostatics they have shown  that while
one-point techniques measure essentially a local rheology,
two-point correlations determine the bulk rheology. Interestingly,
these observations suggest that combined one- and two-point
measurements could be a powerful tool to probe the nature of the
specific interactions between particle and medium.

In this paper, we present an extension of the two-point
correlation technique to high frequencies using optical
interferometry to track particles with high spatial
and temporal resolution. Our approach allows access to a
much faster range of timescales than previous measurements
\cite{1572},
 as we show later. In consequence we are able to measure the rheological
 properties in a frequency range from typically $\sim$ 10$^{1}$
to \unit[10$^{4}$]{rad s$^{-1}$}. 
Accurate measurements of multi-point correlations are
substantially more demanding than one-point measurements.
To achieve adequate signal-to-noise levels we have been
forced to use a more intense laser beam than usually
employed in particle tracking experiments
\cite{1169,1170,1837}.  We consider explicitly the effects
of optical-gradient forces and assume that each particle is
held within a fixed harmonic potential. The two-particle
correlations measured therefore differ from those discussed
by Crocker \textit{et al.} \cite{1572} where diffusion
occurs in the absence of a potential. We develop a
generalised Langevin theory to account for the correlated
time-dependent fluctuations in terms of the viscoelasticity
of the medium. Our paper is organised as follows: In
Section~\ref{sec:theory}, we present an analysis of our
two-particle microrheology experiments, and discuss how
two-particle correlations are related to the
viscoelasticity of the medium (Sec.~\ref{sec:rheology}).
Next, in Section~\ref{sec:materials} we describe our
experimental setup and numerical procedures. Finally in
Section~\ref{sec:results} we present measurements of one-
and two-point fluctuations of two widely-spaced spheres
immersed in a semi-dilute polymer solution which
demonstrate the validity of our approach before we conclude
in Section~\ref{sec:conclude}.

\section{Theory}
\label{sec:theory}

The fundamental concept underlying optical microrheology
experiments is that the Brownian motion of a rigid
spherical tracer particle is
determined by the viscoelasticity of the surrounding
medium. Typically the mean-squared displacement (MSD)
$\left < \Delta x^{2}(t) \right >$ = $\left <
[x(t)-x(0)]^{2} \right
>$ of the particle trajectory is measured (here $x(t)$
denotes the tracer particle position after a time $t$ and
the brackets denote a time or ensemble average). The MSD is
related to the rheological properties of the medium though
linear response theory.
In general, the time-dependent frictional force $F(t)$ on a
sphere moving with a velocity $u(t)$ through a
viscoelastic medium may be written as \cite{1415}
\begin{equation}\label{eqn:linear}
F(t) = \int_{-\infty}^{t} \xi(t-t')u(t') dt',
\end{equation}
where $\xi(t)$ is the instantaneous friction coefficient.
The convolution on the \textit{rhs} of Eq.~\ref{eqn:linear}
reflects the viscoelastic nature of the suspending fluid.
Energy is stored as the Brownian particle diffuses,
generating a ``memory" of the particle's past motion. In
consequence, the frictional force experienced at a time $t$
depends on the particle velocity at all earlier times $t'$.
This non-local time-dependence changes profoundly the
Brownian trajectory so that from a measurement of the MSD
the time-dependent friction coefficient $\xi(t)$ can be
determined. The utility of microrheological measurements
depend on extracting information about the rheological
properties of the medium from the  time dependence of
$\xi(t)$.

\subsection{Correlated Brownian Motion in a Viscoelastic Fluid}
\label{sec:brown}

The optical gradient forces on a high refractive-index
particle within a tightly-focused laser beam are well
approximated by a harmonic interaction. Consequently, we
model our experiments by analysing the dynamics of a pair
of rigid Brownian particles of radius $a$, each held in a
harmonic potential well, and  separated by a distance $r$
within a linear viscoelastic medium. Fluctuations in the
positions of the two particles are correlated as a result
of the hydrodynamic interactions which are transmitted
through the viscoelastic matrix. The standard Langevin
description (see for instance Refs.~\cite{Doi-646,2065})
for the Brownian motion of a pair of neutrally-buoyant
particles of mass $m$ is readily modified to include the
effects of the viscoelasticity of the medium,
\begin{eqnarray}\label{eqn:langevin}
m \frac{du_{1}(t)}{dt} & = &  -\int_{-\infty}^{t}
\xi_{11}(t-t')u_{1}(t') dt' - \int_{-\infty}^{t}
\xi_{12}(t-t')u_{2}(t') dt'   - kx_{1}(t) + f_{1}^{R}(t)
\nonumber \\ m \frac{du_{2}(t)}{dt} & = &
-\int_{-\infty}^{t} \xi_{22}(t-t')u_{2}(t') dt'
-\int_{-\infty}^{t} \xi_{21}(t-t')u_{1}(t') dt'  -
kx_{2}(t) + f_{2}^{R}(t).
\end{eqnarray}

Here the subscripted indices label the two Brownian
spheres, $k$ is the harmonic force constant of the optical
traps (assumed identical), and $f_{i}^{R}(t)$ denotes the
random Brownian forces acting on  particle $i$ ($i$ = 1,2).
The time-dependent frictional coefficients $\xi_{ij}(t)$ in
Eq.~\ref{eqn:langevin} result from the generalisation of
the viscoelastic ``memory" (Eq.~\ref{eqn:linear}) to the
case of two interacting particles. The self term
$\xi_{11}(t)$ details the force acting on one sphere when
the same sphere is moving (and the second sphere is
stationary) while the cross friction coefficient
$\xi_{12}(t)$ describes the force generated on one sphere
by the motion \textit{alone} of the other sphere. Since the
random thermal and frictional forces both have the same
microscopic origin they must satisfy in equilibrium at a
temperature $T$, the fluctuation-dissipation theorem $\left
< f_{i}^{R}(t) f_{j}^{R}(t')\right >$ = $k_{B} T
\xi_{ij}(t-t')$ \cite{1908}.

The coupled generalized Langevin equation
(\ref{eqn:langevin}) is simplified by introducing
collective and relative normal coordinates and using
standard techniques to solve the resulting decoupled
equations of motion. Here we focus on the solution for the
two-particle mean-squared displacement, namely
\begin{equation}\label{eqn:msd}
  d_{ij}(t) = \frac{\left < \Delta x_{i}(t) \Delta x_{j}(t)
  \right >}{\sqrt{\left < \Delta x_{i}^{2} \right > \left < \Delta x_{j}^{2} \right
  >}},
\end{equation}
where the displacements are normalised by the equipartition
result, $\left < \Delta x_{i}^{2} \right
> = 2k_{B}T /k$. The mutual two-particle mean-squared
displacement $d_{ij}(t)$ for $i \neq j$ details the
correlation between the thermal fluctuations of the two
separated particles. The self term $d_{ii}(t)$, is the MSD
of particle $i$ in the presence of a second particle. The
exact solution for the two-particle displacement function
is from Eq.~\ref{eqn:langevin},
\begin{equation}\label{eqn:d12}
  \tilde{d}_{ij}(s) = \frac{k}{2} \left [ \frac{1}{s^{3}m+ s^{2}\tilde{\xi}_{+}
  + sk} + \frac{2\delta_{ij}-1}{s^{3}m+ s^{2}\tilde{\xi}_{-}
  + sk} \right ],
\end{equation}
where the tilde denotes a Laplace transform, $\tilde{f}(s)$
= $\int_{0}^{\infty} f(t) \textrm{e}^{-st} dt$, $s$ is the
Laplace variable and $\tilde{\xi}_{\pm}$ =
$\tilde{\xi}_{11} \pm \tilde{\xi}_{12}$. The complete
solution (\ref{eqn:d12}) is still too complicated to use to
analyse experimental data. In order to derive a simpler
result we make two simplifying assumptions.

We start by considering the importance of particle inertia.
The mass $m$ of the particle determines the decay time
$\tau_{B} \sim m / 6 \pi \eta_{\infty}a$ of the particle
velocity autocorrelation function. Here $\eta_{\infty}$ is
the high frequency viscosity of the medium. For frequencies
small compared to $\omega_{B} = 1/\tau_{B}$ the momentum of
the Brownian particle will have relaxed to zero and we may
neglect particle inertia. In our experiments, $\omega_{B}$
is of order  \unit[10]{MHz} so neglecting particle inertia
is an excellent approximation at frequencies of
experimental interest ($\omega \leq $ \unit[20]{kHz}).

Secondly, we consider only the situation where the two probe
spheres are separated by a distance $r$ which is large in
comparison to the sphere radius $a$. In this limit, Batchelor
\cite{Batchelor-618} has argued from general symmetry
considerations that the hydrodynamic functions depend on the
dimensionless sphere separation $\rho$ = $r/a$ as,
\begin{eqnarray}\label{eqn:friction}
\xi_{11} & = & a_{0} + \frac{a_{2}}{\rho^{2}} +
\frac{a_{4}}{\rho^{4}} + \mathcal{O}(\rho^{-6}) \nonumber
\\
\xi_{12} & = & - \frac{b_{1}}{\rho} - \frac{b_{3}}{
\rho^{3}} - \mathcal{O}(\rho^{-5}),
\end{eqnarray}
where the coefficients $a_{i}$ and $b_{i}$ are constants
for a given medium. This asymptotic solution reveals that
provided the two spheres are far enough apart, the self
friction $\xi_{ii}$ is, to leading order, independent of
the separation $\rho$. The mutual friction coefficient
$\xi_{12}$ by contrast depends sensitively on the pair
separation varying as $1/\rho$ at large $r$ so that the
ratio $ \xi_{12} / \xi_{11}$ is a small parameter which
shrinks with increasing separation as $1/\rho$.
Consequently the exact expression  for the two-particle
displacements (Eq.~\ref{eqn:d12}) may be simplified by
expanding $\tilde{d}_{ij}$ in powers of $ \tilde{\xi}_{12}
/ \tilde{\xi}_{11}$. At large $\rho$, only the constant
terms and the part which decays with distance as $1/\rho$
contribute to the displacement field. Retaining these
leading terms yields the desired simplified expression for
the correlated sphere displacements. Ignoring particle
inertia, the one- and two-particle displacements may then
be expressed as
\begin{eqnarray}\label{eqn:simple}
\tilde{d}_{ii}(s) & = & \frac{k}{s(s\tilde{\xi}_{ii}+k)}
\nonumber
\\
 \tilde{d}_{ij}(s) & = & -  \frac{k
 \tilde{\xi}_{ij}}{(s\tilde{\xi}_{ii}+k)^{2}},
\end{eqnarray}
where $i \neq j$. Rearranging these equations gives
expressions for the friction coefficients in terms of
experimental quantities, namely
\begin{eqnarray}\label{eqn:solution}
\frac{s}{k} \tilde{\xi}_{ii}(s) & = & \frac{1}{s
\tilde{d}_{ii}(s)} - 1 \nonumber
\\
& & \nonumber \\
 \frac{s}{k} \tilde{\xi}_{ij}(s) & = & \frac{-\tilde{d}_{ij}(s)}{s
 \tilde{d}_{ii}^{2}(s)}.
\end{eqnarray}
Measurements of the single-particle fluctuations yield the
self friction $\tilde{\xi}_{ii}(s)$, whereas the mutual
friction coefficient is found from a knowledge of both
single-particle and two-particle positional fluctuations.
Our analysis is valid for any viscoelastic media, we make
no assumptions about the homogeneity of the medium. We
assume only that: (i) the particle pair separation $r$ is
large in comparison to the sphere radius $a$, and (ii)
particle inertia  is unimportant.

\subsection{Relation to Rheology}
\label{sec:rheology}

\begin{figure}
\includegraphics[width=6in]{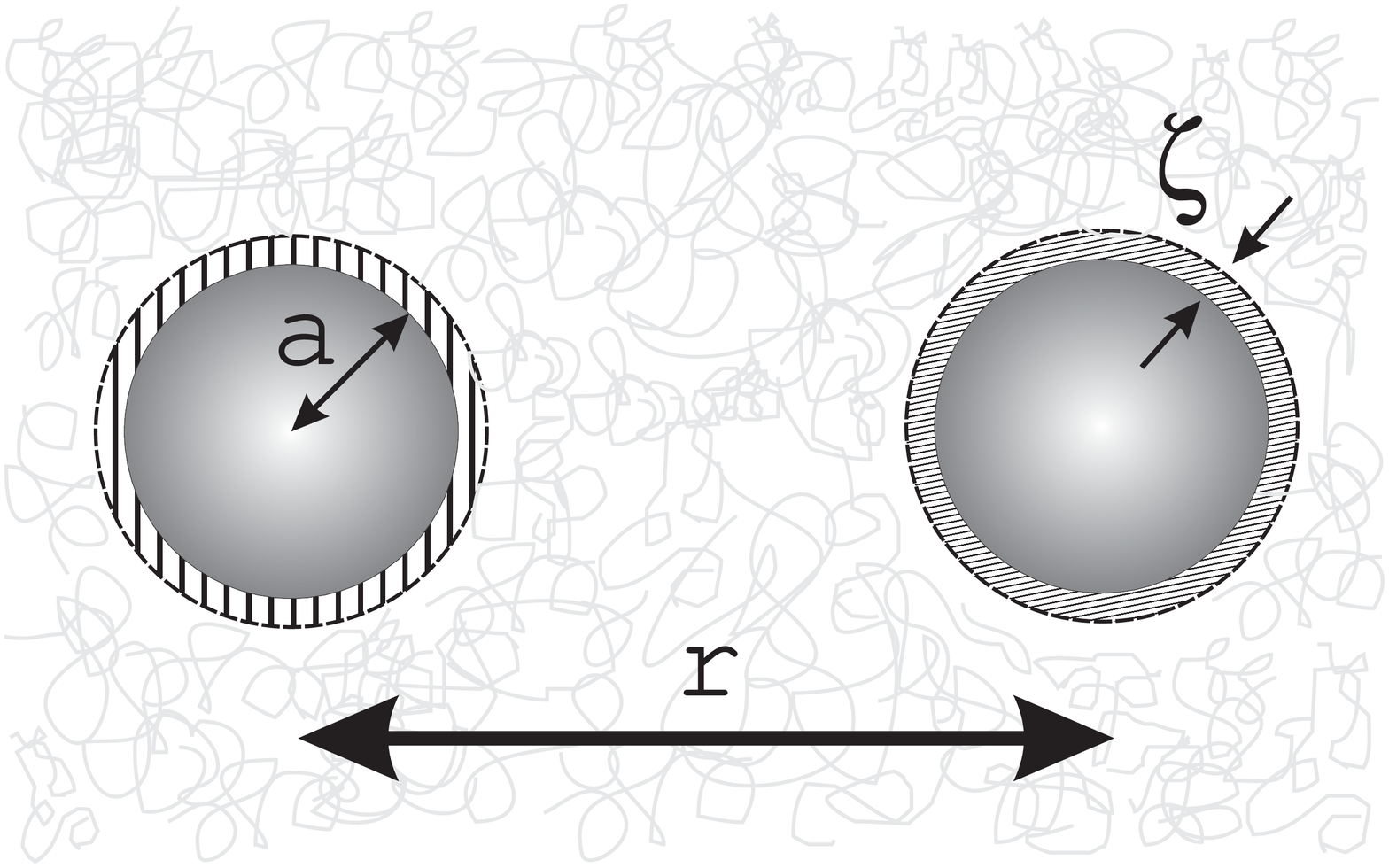}
 \caption{Schematic illustration of two-point microrheology. Two spheres of
 radius $a$ are suspended a distance $r$ apart in a semi-dilute polymer solution. The non-adsorbing polymer is depleted from
 an interfacial  region of width  $\zeta$ surrounding each particle (shown hatched). The dynamics of
 an individual particle is a sensitive function of the width and properties of the interfacial
 zone. By contrast, the correlated \textit{pair} fluctuations depend upon the mechanical properties of the medium,
 averaged on the longer scale $r$. In the limit when $r \gg a \gg \zeta$ the pair fluctuations measure the bulk
 rheology.}
 \label{fig:micro}
 \end{figure}

The relationship between  the friction coefficient
$\tilde{\xi}_{ij}(s)$ and the viscoelasticity of the medium is a
challenging problem in fluid mechanics. Some of the complications
are illustrated in Figure~\ref{fig:micro} which shows
schematically the microstructure of a colloid-polymer mixture. If
the polymer is non-adsorbing then around each particle is a
depletion zone of width $\zeta$ where the polymer concentration is
less than in the bulk. Physically, we  expect the motion of an
individual sphere to depend on the \textit{local} environment
rather than the rheology of the bulk solution. So it seems
reasonable to expect the depletion zone around a particle to speed
up diffusion, in comparison to the case where the medium is
homogeneous.

While such specific interactions between  probe and medium
have not been considered, Levine and Lubensky \cite{1547}
have analysed in detail the simpler case of a viscoelastic
continuum. Over a wide range of frequencies, they predict
that the single particle friction coefficient satisfies the
generalised Stokes-Einstein relation,
\begin{equation}\label{eqn:GSER}
\tilde{\xi}_{ii}(s) = \frac{6 \pi a \tilde{G}(s)}{s},
\end{equation}
where $\tilde{G}(s)$ is the Laplace shear modulus of the
homogeneous medium. This result reduces to the well-known
Stokes expression ($\xi$ = $6 \pi \eta a$) for the friction
of a rigid sphere in a viscous fluid \cite{1415} if the
Laplace modulus is replaced by its Newtonian limit,
$\tilde{G}(s) = s \eta$. Deviations from Eq.~\ref{eqn:GSER}
are expected if the local environment around each sphere is
not the same as the bulk.

While probe--particle interactions  have a significant effect on
single particle motion, Crocker \textit{et al.} \cite{1572} have
suggested that they should have a much weaker influence on the
correlated fluctuations of \textit{pairs} of particles. This
conjecture has recently been confirmed theoretically \cite{1547}.
In a homogeneous viscoelastic medium, the two-particle friction
coefficient has the form,
\begin{equation}\label{eqn:mutual}
\tilde{\xi}_{ij}(s) = - \frac{9 \pi a \tilde{G}(s)}{\rho
s}.
\end{equation}
This equation is valid provided the pair separation $r$ is large
in comparison to the sphere radius $a$. In the Newtonian limit, we
recover the asymptotic form of the mutual friction coefficient for
a viscous fluid ($\xi_{ij}$ = $- 9 \pi \eta a / \rho$
\cite{Batchelor-618}).

\section{Materials and Methods}
\label{sec:materials}

\subsection{Dual-beam Optical Tweezers}
\label{sec:tweezers}

\begin{figure}
\includegraphics[width=6in]{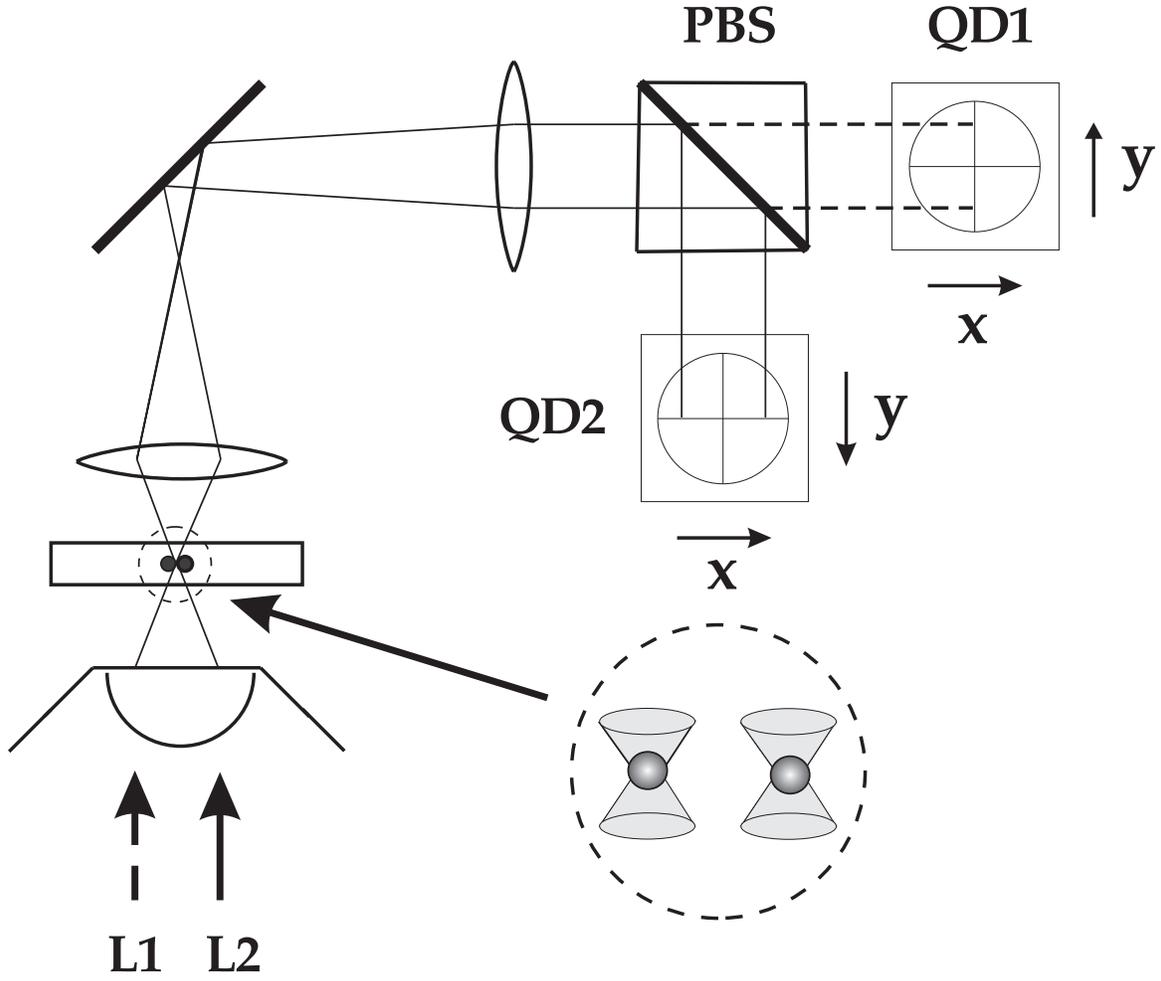}
 \caption{Dual-beam optical tweezer setup: Optical traps are generated by
focusing two orthogonally-polarized IR laser beams
 (L1 and L2) using a high numerical aperture microscope objective. Intensity shifts caused by interference between
 the direct beam and light scattered by the trapped sphere are imaged onto a pair of quadrant detectors (QD1 and QD2).
  A polarizing
 beam splitter (PBS) is used to separate the two orthogonal signals. The $x$ and $y$ positions are derived
 by combining the voltages from the four segments of the quadrant photodiodes using low-noise analog electronics.
 The signals are then digitized and stored on a PC.}
 \label{fig:expt}
 \end{figure}

The details of the dual-beam optical tweezer equipment used
in the present experiments have been described in a
preceding publication \cite{2065}. Figure~\ref{fig:expt}
shows schematically the apparatus. In brief, two spherical
colloidal particles are held by optical gradient forces
near the focus of a pair of orthogonally-polarised laser
beams ($\lambda$ = \unit[1064]{nm}). The mean separation
$r$ between the probe particles was varied by altering the
positions of the trapping lasers with an external
computer-controlled mirror. The thermal fluctuations in the
position of each probe particle were monitored by recording
the interference between the forward scattered and
transmitted infra-red beams with a pair of quadrant
photodetectors. Custom-built current-to-voltage converters
allowed the in-plane positions of both spheres to be
recorded with a spatial resolution of $\sim$ \unit[1]{nm}
at time intervals of \unit[50]{$\mu$s}.

The optical gradient force on a sphere displaced from the focus of
the laser beam is harmonic. The force constant $k$ of each trap
was determined in a separate calibration experiment \cite{2065}.
Colloidal spheres dispersed in decalin were trapped in each of the
two beams in turn. The single particle mean squared displacement
$\left < \Delta x^{2}(t) \right >$ was measured and fitted to
$\langle \Delta x^{2}(t) \rangle = \langle \Delta x^{2} \rangle
\left[1-\exp(-t/\tau) \right]$ to yield the trap stiffness $k$ =
$6 \pi \eta a / \tau$ from known values of the particle radius $a$
and solvent viscosity $\eta$. The intensity of the two beams was
adjusted until the stiffness of the traps differed by less than
$\sim$ 5\%. For a given geometry and laser intensity, the harmonic
force constant $k$ is a function only of the refractive index
mismatch between particle and medium, so these calibration
constants are expected to hold for the polymer data presented
below. The root-mean-square fluctuation in the particle position
in each trap was \textit{ca.} \unit[50]{nm}.

\subsection{Numerical Methods}

The normalised one- and two-particle mean-squared
displacements $d_{ij}(t)$ were calculated from the measured
particle trajectories $\{x_{1}(t),x_{2}(t)\}$ using fast
Fourier transform algorithms \cite{2065}. The Laplace
transform of the one-particle MSD $d_{ii}(t)$ was
determined from a regularised fit to a linear superposition
of exponential terms,
\begin{equation}\label{eqn:LT:d11}
  d_{ii}(t) = 1 - \sum_{k=1}^{N} L_{k}
  \textrm{e}^{-t/\tau_{k}},
\end{equation}
where $N$ is the number of terms, $\tau_{k}$ is the decay
time specified on a logarithmic grid and the coefficients
$L_{k}$ were determined by requiring that the spectrum of
relaxation times remain smooth and penalizing sums
proportional to their second derivative \cite{2331}. The
curvature penalty suppresses unphysical wild oscillations
in the solution.  The constrained regularised solution to
Eq.~\ref{eqn:LT:d11} was found using the CONTIN algorithm
\cite{2332}. Once $L_{k}$ have been determined the Laplace
transform may be calculated exactly as,
\begin{equation}\label{eqn:d11s}
 \tilde{d}_{ii}(s) = \frac{1}{s} - \sum_{k=1}^{N} L_{k}
 \frac{1}{s+\tau_{k}^{-1}}.
\end{equation}

The Laplace transform of the two-particle MSD $d_{ij}(t)$
was evaluated by a two step numerical procedure. First,
$d_{ij}(t)$ was fitted to a series of cubic polynomial
splines. The contribution to the transform from each time
interval was then expressed in terms of a series of
incomplete gamma functions which were evaluated
numerically. Tests showed that errors in the transform were
of the order of 5\%, except near the frequency extremes
where truncation errors became more significant.

\section{Results}
\label{sec:results}

To test our approach we have made measurements on a
viscoelastic semi-dilute solution  of polystyrene (PS) with
an average molecular weight of 10$^{7}$
(M$_{\textrm{W}}$/M$_{\textrm{N}}$ = 1.19) in a mixed
\textit{cis}- and \textit{trans}-decalin solvent (volume
fraction \textit{cis}-decalin 0.48).  Linear polystyrene in
decalin is a well studied system \cite{133}. The
$\Theta$-temperature of PS in an equal volume mixture of
\textit{cis}- and \textit{trans}-decalin is \unit[$\sim$
16]{\degrc} so at the temperature of our measurements
(\unit[$\sim$ 23]{\degrc}) decalin remains  a near-$\Theta$
solvent. Using literature values \cite{133} we estimate the
polymer radius of gyration at \unit[23]{\degrc} as
$R_{g}\sim $ \unit[102]{nm}, slightly larger than the value
under $\Theta$-conditions of \unit[87]{nm}. The overlap
concentration $c^{*}$ is \unit[3.7]{mg cm$^{-3}$}.
Sterically-stabilised poly(methyl methacrylate) spheres of
radius $a$ = \unit[0.643]{$\mu$m} were added as probe
particles at a volume fraction $\phi \approx$ 10$^{-6}$.
The viscous suspensions were loaded into flat, rectangular
capillary tubes \unit[170]{$\mu$m} thick which were
hermetically sealed with a fast-setting epoxy glue and
mounted onto a microscope slide. The samples were left for
at least forty minutes to allow the particles to settle to
the bottom surface before the beginning of an experiment.
The trajectories of two spheres were collected at ten
roughly even-spaced pair separations between
\unit[3]{$\mu$m} and \unit[17]{$\mu$m}. 2$^{23}$ data
points were collected at each separation at a sampling
frequency of \unit[20]{kHz}. To avoid wall effects, we
analysed only spheres at least \unit[15] {$\mu$m} away from
the capillary walls. All measurements were performed at
room temperature.

\begin{figure}
\leavevmode
\begin{minipage}[b]{8cm}
\begin{center}
\includegraphics[width=8cm]{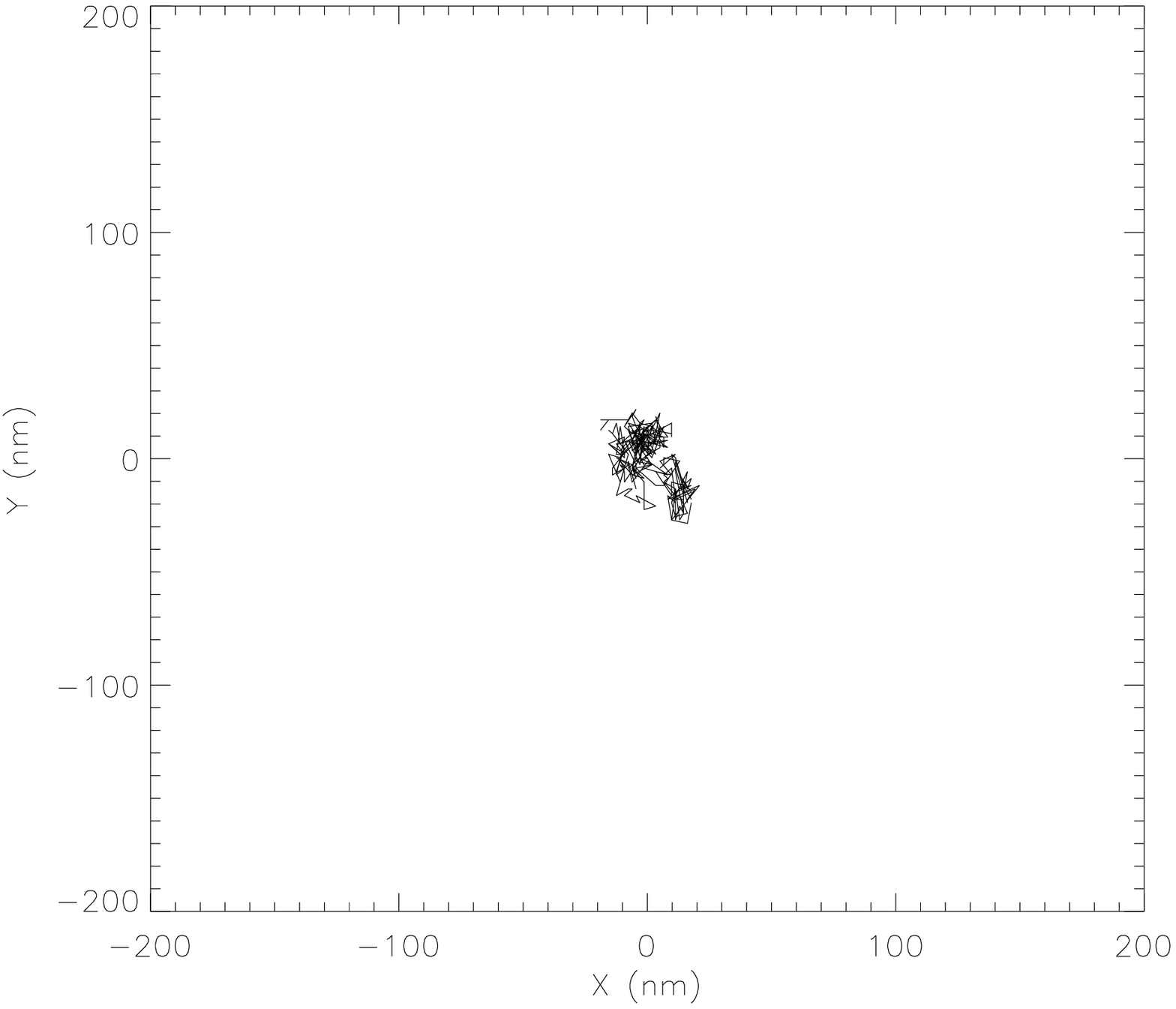}
\\  (a)
\end{center}
\end{minipage}
\hfill
\begin{minipage}[b]{8cm}
\begin{center}
\includegraphics[width=8cm]{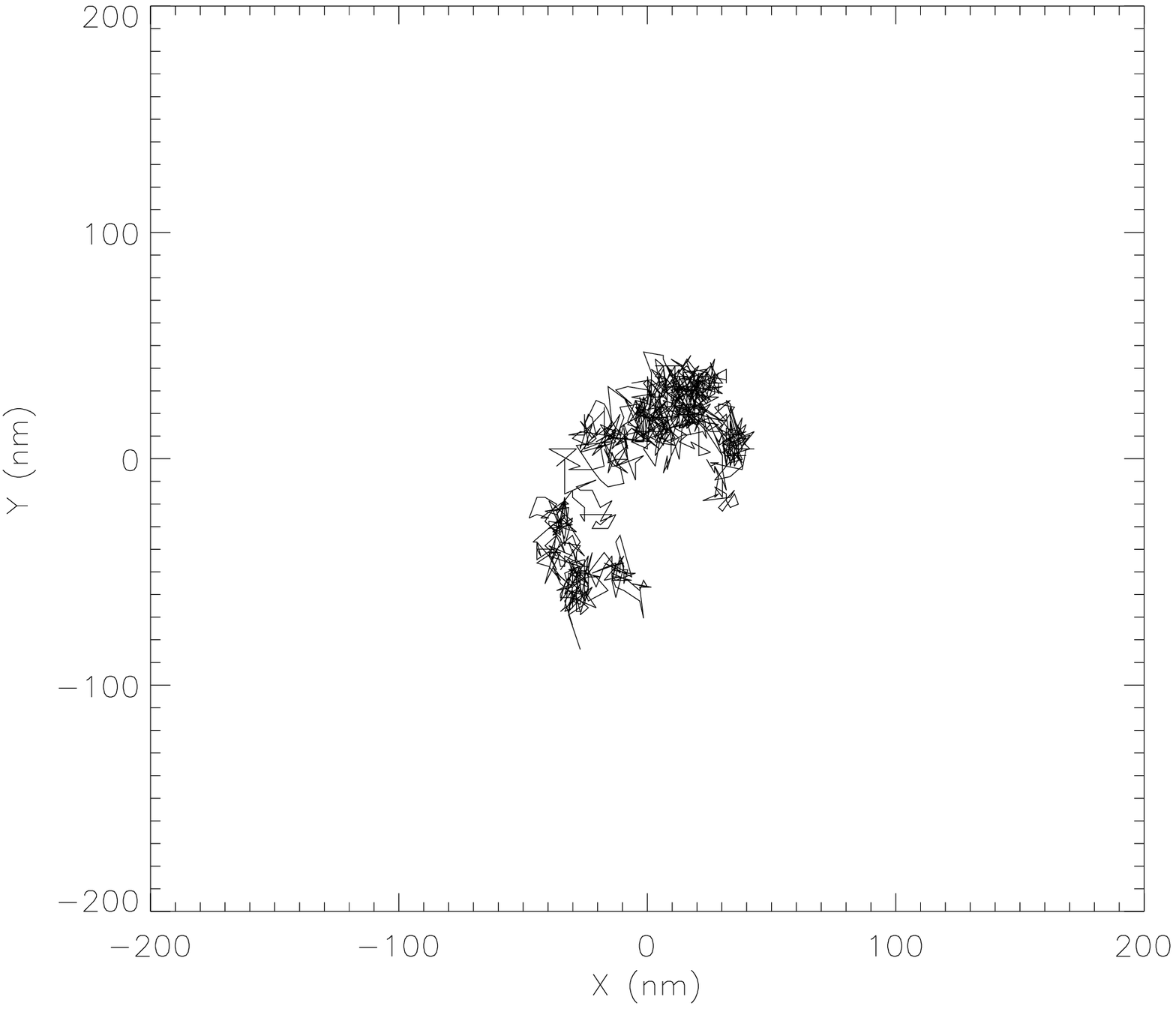}
\\(b)
\end{center}
\end{minipage}
\hfill
\begin{minipage}[b]{8cm}
\begin{center}
\includegraphics[width=8cm]{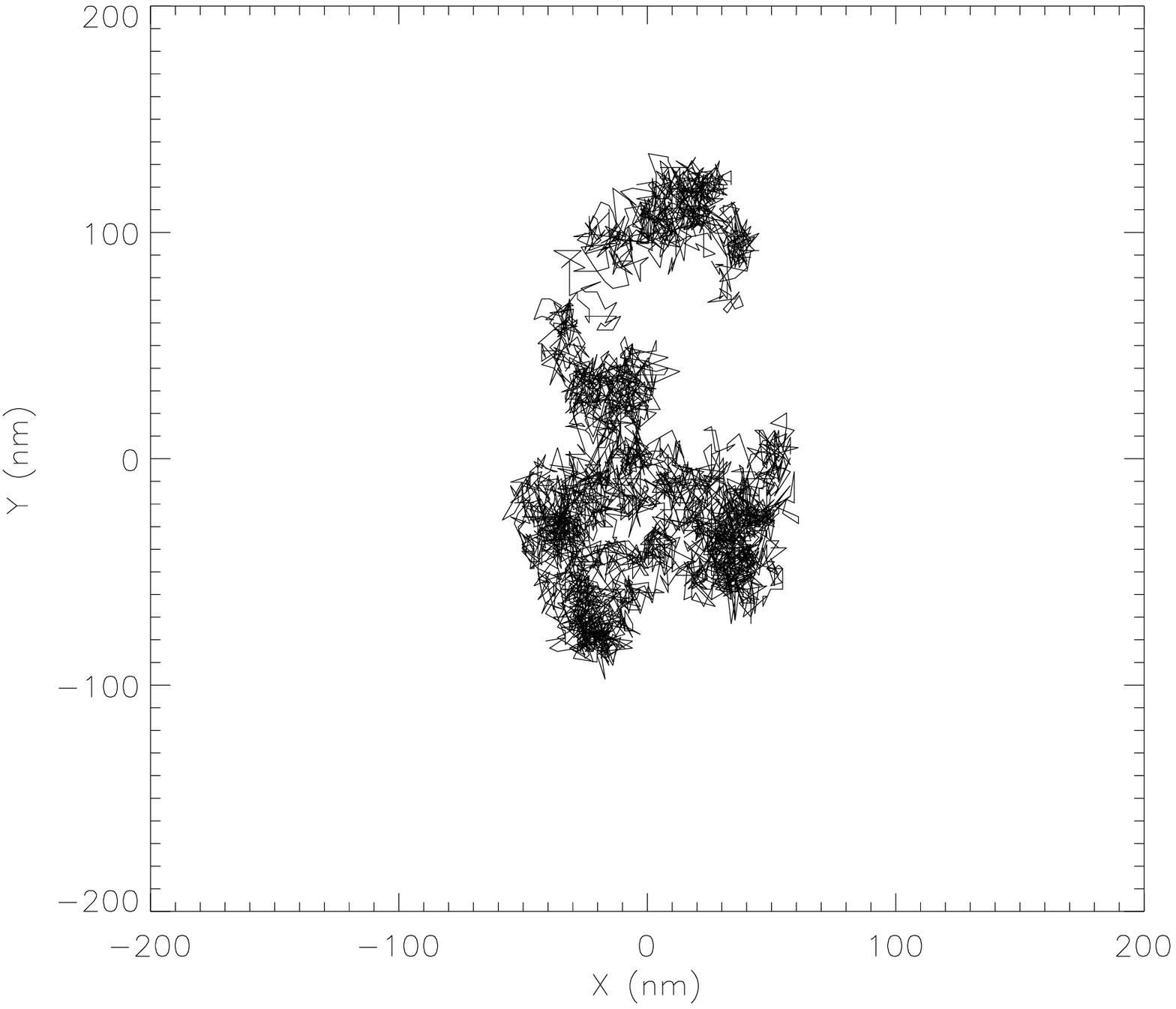}
\\(c)
\end{center}
\end{minipage}
\hfill
\begin{minipage}[b]{8cm}
\begin{center}
\includegraphics[width=8cm]{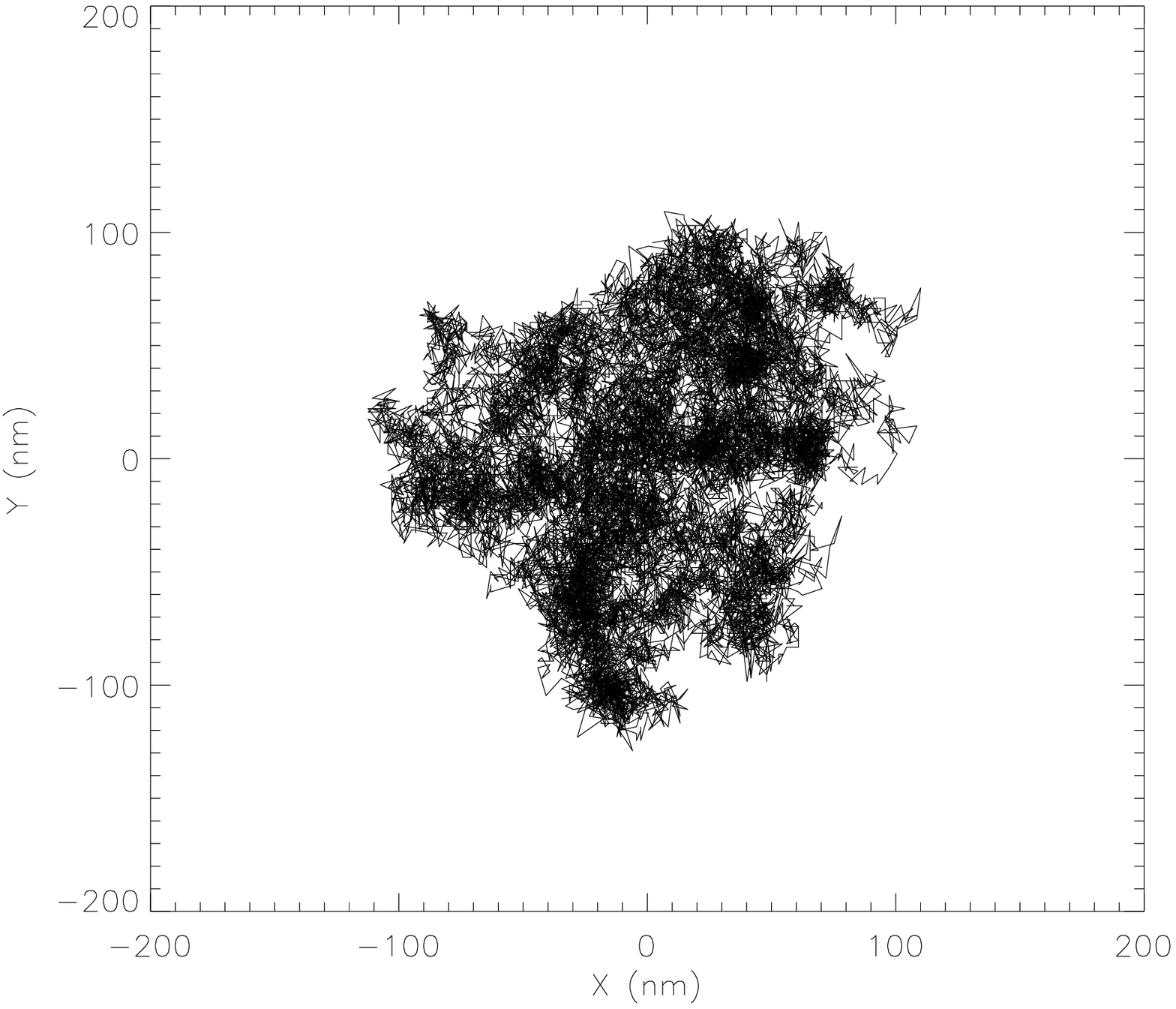}
\\(d)
\end{center}
\end{minipage}
\caption{The measured two-dimensional trajectory of one of
two widely separated spheres in a semi-dilute PS solution
($c$ = \unit[1.7]{$c^{*}$}) over a duration of (a)
\unit[0.0128]{s}, (b) \unit[0.0512]{s}, (c) \unit[0.2048]
{s}, and (d) \unit[0.8192]{s}. The sampling frequency is
\unit[20]{kHz}.} \label{fig:traj}
\end{figure}

The time-dependent trajectory of a sphere diffusing in a PS
solution  is illustrated in Fig.~\ref{fig:traj}. In
contrast to cross-linked gels, the entanglements present in
a semi-dilute polymer solution are only temporary so on a
sufficiently long time scale the solution is expected to be
purely viscous. Consequently the trajectory, when averaged
over a long enough period, will depend only on the external
potential and not on the viscoelasticity of the solution.
The optical forces are harmonic so that at long times the
particle probability distribution will be Gaussian and the
mean squared displacement will plateau at the equipartition
limit, $\left < \Delta x^{2}(t) \right
>$ = $2 k_{B} T / k$.
This behaviour is seen in Fig.~\ref{fig:traj}~(d). At short
times, the particle statistics are also Gaussian although
diffusion is now hindered by the temporary entanglements
found in the surrounding semi-dilute polymer solution
(Fig.~\ref{fig:traj}~(a)). The non-Gaussian dynamics at
intermediate times (Fig.~\ref{fig:traj}~(b) and (c)) are
most intriguing. The trajectory is highly anisotropic and
does not resemble the isotropic diffusion seen on either
shorter or longer time intervals. The particle seems to
make infrequent large ``jumps'' over a distance which is
comparable to the polymer radius of gyration. This
behaviour suggests that the instantaneous micro-environment
around each sphere in a polymer solution is heterogeneous
and evolves in time.

\begin{figure}
\includegraphics[width=6in]{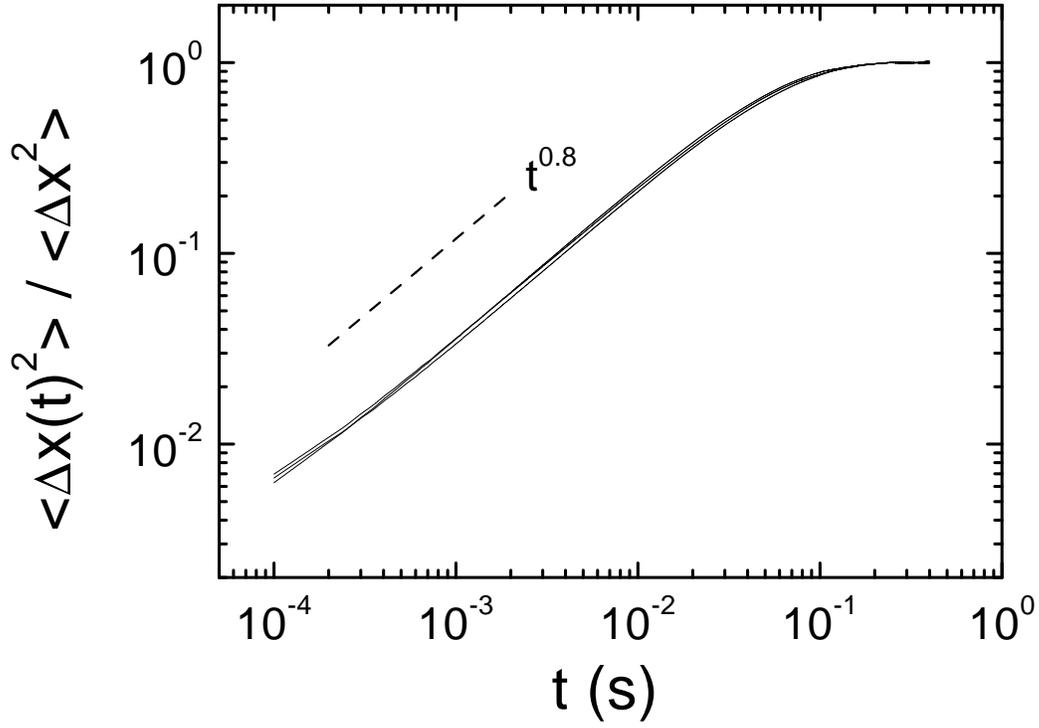}
 \caption{Single particle mean squared displacements  measured for a pair of widely-separated
 particles ($c$ = \unit[1.7]{$c^{*}$}). The different curves depict data collected at three different pair separations ranging
 from $\rho$ = 4.53 to 16.96. No systematic variation with $\rho$ is evident.}
 \label{fig:d11}
 \end{figure}

To characterise  the mechanical environment in detail we
calculated the mean-squared $x$-displacement, $\left < \Delta
x^{2}(t) \right >$, from the trajectory of a single particle. The
data was recorded from a \textit{pair} of widely-separated spheres
trapped within a PS solution. Fig.~\ref{fig:d11} depicts one of
the two MSDs, normalised by the plateau displacement, at a number
of different pair separations. Data for the second sphere was
essentially identical and is not shown. As expected, there is no
systematic dependence of the single particle MSD on the pair
separation $r$, provided the two spheres remain well separated ($r
\gg a$). The measured $\left < \Delta x^{2}(t) \right >$ is
sensitive only to the local rheological environment around each
particle. Because of the high frequency elasticity of the polymer
solution, $\left < \Delta x^{2}(t) \right >$ increases
approximately as $t^{0.8}$ at early times rather than linearly, as
expected for diffusion in a purely viscous medium. At long times
$\left < \Delta x^{2}(t) \right >$ reaches the equipartition
plateau.

\begin{figure}

\includegraphics[width=6in]{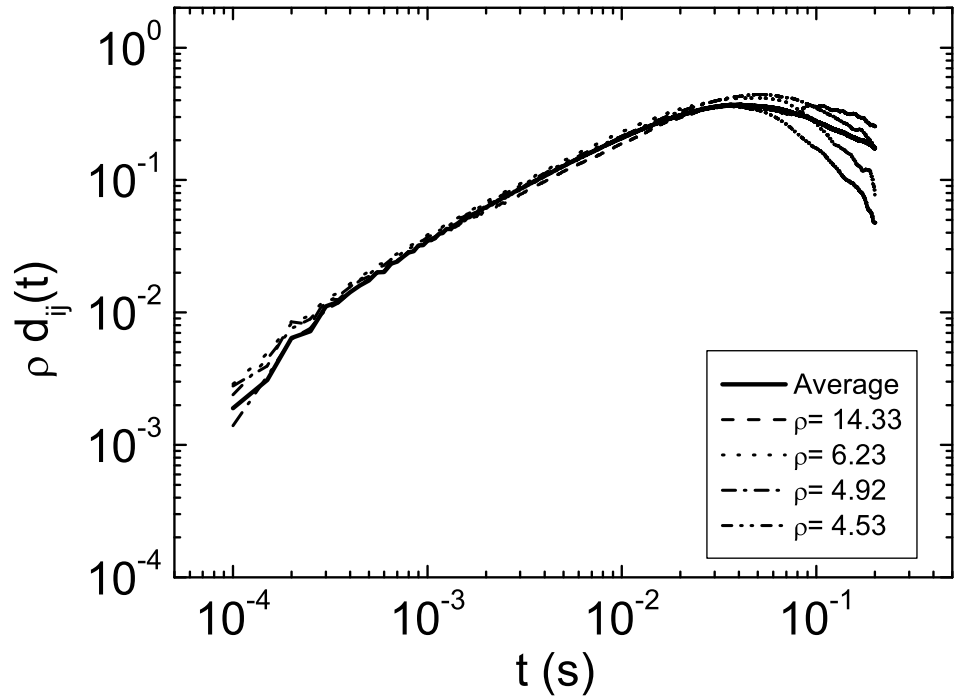}
 \caption{The time dependence of the scaled mutual two-particle mean squared displacement $\rho
 d_{ij}(t)$ measured at four different pair separations
 $\rho$ ($c$ = \unit[1.7]{$c^{*}$}). For intermediate times all of the data collapse
 onto a common curve showing that the strength of the two particle
 correlations varies inversely with the pair spacing $\rho$.
 The solid line is the average for all of the data collected.}
 \label{fig:d12}
 \end{figure}

Using the same trajectory data, we have also calculated the
two-point correlations. We expect that for widely-spaced
particles the strength of the correlated positional
fluctuations will decay inversely with the particle
spacing. To confirm this, we plot in Fig.~\ref{fig:d12} the
mutual mean-squared displacement $d_{ij}(t)$ multiplied by
the dimensionless spacing $\rho$. The data measured for a
range of different pair spacings is seen to collapse onto a
common curve over a wide range of times, although
deviations are evident at very long times. The variations
seen at long times do not depend systematically on the pair
spacing  $\rho$ and are probably a reflection of poor
statistics or mechanical vibrations. The peak seen in
$d_{ij}(t)$ is a consequence of the applied optical
potential since for very long times the correlations
between the two particles must decay ultimately back to
zero.

\begin{figure}

\includegraphics[width=6in]{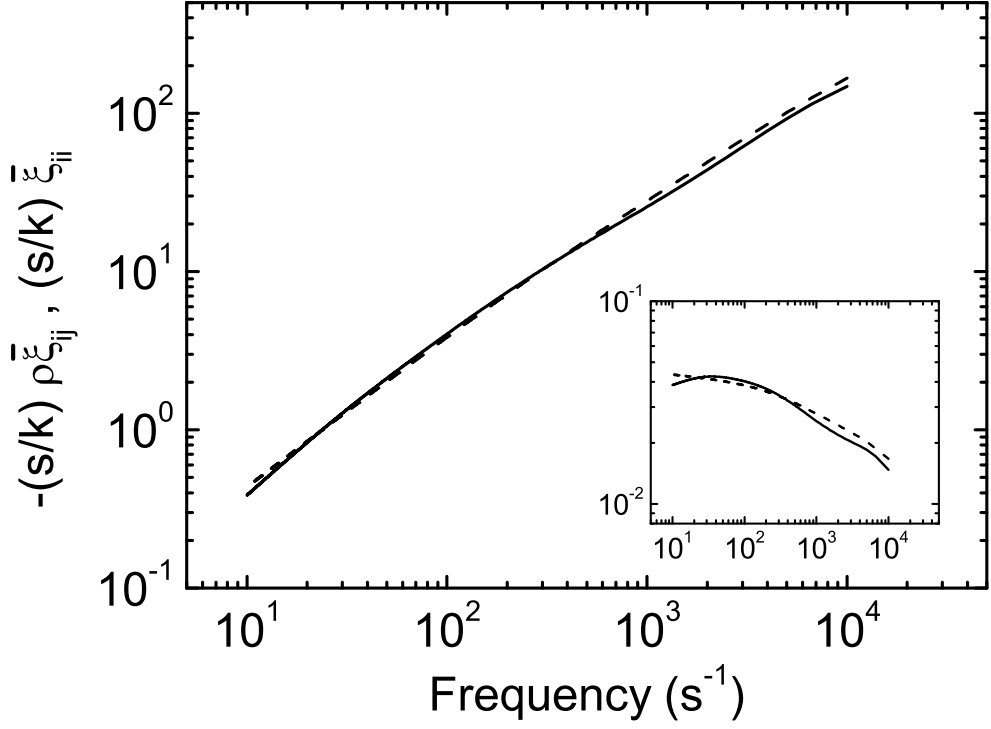}
 \caption{Comparison of the frequency dependence of the scaled two-particle $-(s/k) \rho \tilde{\xi}_{ij}(s)$ (solid)
 and single-particle friction coefficients $(s/k) \tilde{\xi}_{ii}(s)$ (dashed) determined from the correlated fluctuations
 of two widely-spaced particles in a semi-dilute ($c$ = \unit[1.7]{$c^{*}$}) polymer solution. The inset diagram shows $-(\rho/k) \tilde{\xi}_{ij}(s)$ (solid)
 and $\tilde{\xi}_{ii}(s)/k$ (dashed) as a function of frequency.}
 \label{fig:friction}
 \end{figure}

The one- and two-particle mean-squared displacements contain
features which originate from the viscoelasticity of the polymer
solution and from the applied optical forces. To separate the two
contributions we use the analysis of Sec.~\ref{sec:theory}.
Inverting $d_{ii}(t)$ and $d_{ij}(t)$ yields friction coefficients
which, apart from a simple scaling, do not depend on the applied
potential. The computed friction coefficients  are shown in the
inset in Fig.~\ref{fig:friction}. The frequency dependence seen is
in sharp contrast to that expected for a purely viscous fluid
where $\tilde{\xi}_{ij}$ remains independent of frequency. The
steady decrease in the particle friction with increasing frequency
reflects the viscoelasticity of the polymer solution and in
particular its ability to store energy elastically at high
frequencies. To discuss further the connection between the
particle friction and the medium rheology we focus on the
frequency scaled functions $s\tilde{\xi}_{ii}$ and $-s \rho
\tilde{\xi}_{ij}$. In the continuum model of Levine and Lubensky
\cite{1547,2009} both functions are predicted to be simple
multiples of the Laplace-transformed shear modulus $\tilde{G}(s)$
so that the ratio $-s \rho \tilde{\xi}_{ij} / s\tilde{\xi}_{ii}$
should be frequency-independent and from, Eqs.~\ref{eqn:GSER} and
\ref{eqn:mutual}, equal to $3/2$. The experimentally-derived
values for $s\tilde{\xi}_{ii}$ and $-s \rho \tilde{\xi}_{ij}$ are
shown in Fig.~\ref{fig:friction}. In partial accord with these
predictions, we find that the two scaled friction coefficients do
exhibit a very similar functional form (except possibly at the
highest frequencies). However the ratio $-s \rho \tilde{\xi}_{ij}
/ s\tilde{\xi}_{ii}$ is not $3/2$ but of order unity.

While we do not have a complete understanding of this discrepancy
one possible explanation is that the nature of the shear coupling
between the particle and the medium differs in the one- and
two-particle situations. In the immediate vicinity of a probe
particle a depletion zone exists with a reduced polymer segment
concentration. While hydrodynamic forces will still couple the
motion of the probe particle to the polymer matrix, the depletion
zone could modify the nature of the boundary conditions at the
particle surface. For a rigid spherical particle, the choice of
boundary conditions changes the friction coefficient by a factor
of $3/2$ from $6 \pi \eta a$ for stick to $4 \pi \eta a$ for slip
\cite{1415}. Cardinaux \textit{et al.} \cite{2322} have recently
suggested that a slip, as opposed to a stick, boundary condition
might apply to single particle motion in giant-micellar solutions.
Assuming the depletion of polymer has a similar effect, then we
should identify $\tilde{\xi}_{ii}(s)$ with $4 \pi a \tilde{G}(s) /
s$ rather than Eq.~\ref{eqn:GSER}. In the case of slip, the
two-particle friction coefficient $\xi_{ij}$ is  $-\rho
\tilde{\xi}_{ij}(s)$ = $4\pi a \tilde{G}(s) / s$ \cite{jones}. In
this case the ratio $- \rho \tilde{\xi}_{ij} / \tilde{\xi}_{ii}$
is $1.0$ which is in close agreement with the value found
experimentally. Further experimental and theoretical work is
necessary to confirm this picture. A detailed comparison between
the one- and two-particle friction coefficients as a function of
polymer concentration and molecular weight is currently underway
and will be the subject of future publications.

\begin{figure}
\includegraphics[width=6in]{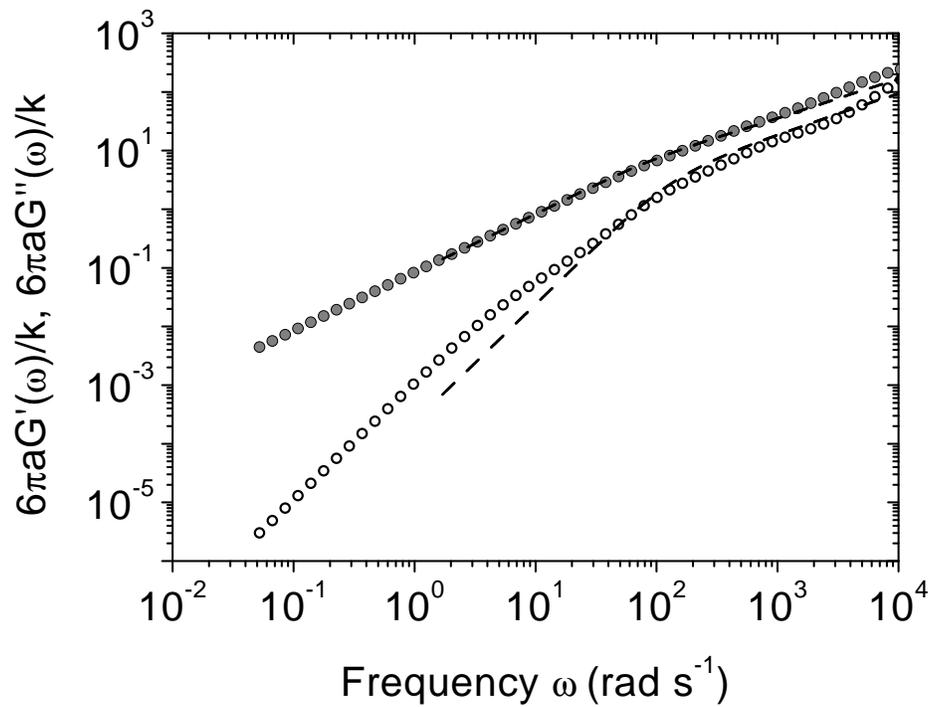}
 \caption{Scaled storage $6 \pi a G'(\omega) /k$ (open circles) and loss moduli $6 \pi a G''(\omega) /k$
 (filled circles) as function of frequency for \unit[2.9]{$c^{*}$}
 solution of polystyrene in decalin. The dashed lines show predictions from Zimm theory.}
 \label{fig:moduli}
 \end{figure}

The frequency-dependent complex shear modulus
$G^{*}(\omega)$ = $G'(\omega) + iG''(\omega)$  may be
determined from either the one- (after allowing for the
slip boundary conditions) or two-point friction
coefficients. Fig.~\ref{fig:moduli} shows the storage and
loss moduli computed from tweezer measurements for a
\unit[2.9]{$c^{*}$} solution of polystyrene. The data
exhibits the features expected for an entangled flexible
polymer solution \cite{Doi-646}.  At high frequencies,
above the maximum relaxation time of the solution, the
shear and loss moduli  are expected to show a power law
dependence on $\omega$. Under $\Theta$--conditions, $G'$
and $G''$ are predicted by the Zimm model to vary as
$\omega^{2/3}$. Fig.~\ref{fig:moduli} shows that  at high
frequencies the complex modulus $G^{*}(\omega)$ follows
closely the predictions of Zimm theory. At low frequencies
the elasticity of the entanglement network contributes and
the measured $G'({\omega})$ is increased above the Zimm
predictions as expected for a semi-dilute solution.

\section{Summary}
\label{sec:conclude}

In this paper we have demonstrated that combined
measurements of one- and two-particle thermal fluctuations
can be used to provide new insights into the the nature of
the interface around a probe sphere embedded in a
viscoelastic medium. We have measured the correlated
Brownian motion of two widely-separated spheres held by a
pair of tightly-focused laser beams within a polymer
solution. Photodetection of scattered light was used to
record the trajectories of the trapped particles with high
spatial and temporal resolution. This new technique differs
from previous two-point microrheological studies
\cite{1572} in its spatial resolution and access to high
frequencies; frequencies up to \unit[10$^{4}$]{rad
s$^{-1}$} are readily detectable. In comparison, the
frequency range of previous two-point measurements
\cite{1572} have been limited by the speed of video
microscopy  to $\omega \sim $ \unit[10$^{2}$]{rad
s$^{-1}$}. Faster A/D data acquisition should enable the
optical tweezer technique to be extended to still higher
frequencies approaching \unit[10$^{7}$]{rad s$^{-1}$}.

In order to achieve good statistics on two-particle
correlations we have used a significantly higher-powered
laser beam than previous (single-particle) optical tracking
studies \cite{1169,1170,1837}. As a result we are no longer
able to ignore the effects of optical forces applied by the
laser beam on the measured particle fluctuations, as
previous studies have assumed implicitly
\cite{1169,1170,1837}. We account for the optical-gradient
forces by analysing the dynamics of two harmonically-bound
particles embedded in a general viscoelastic medium. By
making two simplifying assumptions, which we show are valid
in our experiments, we develop simple expressions for the
frequency-dependent single- and two-particle friction
coefficients, in terms of measurable quantities. We expect
the single-particle friction coefficient to be sensitive to
the rheological properties of the interfacial zone
immediately surrounding a probe sphere while two-point
measurements should allow the bulk rheological properties
to be determined.

We have demonstrated the validity of our approach by analysing the
one- and two-point correlations measured for \unit[1.28]{$\mu$m}
diameter poly(methyl methacrylate) spheres suspended in a
semi-dilute solution of polystrene. One- and two-particle friction
coefficients are computed. We confirm that over a wide range of
particle separations, the two-point correlations vary inversely
with the particle spacing $r$ while the one-point correlations are
independent of $r$. The two friction coefficients show a very
similar dependence on frequency with the single-particle friction
coefficient $\tilde{\xi}_{ii}(s)$ approximately equal to the
corresponding two-particle friction coefficient $-\rho
\tilde{\xi}_{ij}(s)$. While viscoelastic theory confirms that the
two functions should have a similar frequency dependence, the
ratio measured is in quantitative disagreement with recent
theoretical calculations for a viscoelastic continuum \cite{1547},
which predict a ratio of $3/2$. We suggest that this discrepancy
reflects the existence of a depletion zone around each particle in
a non-adsorbing polymer solution. As a consequence, the shear
coupling between particle and medium may be approximated by a slip
rather than a stick boundary conditions on the particle surface.
With this assumption, either the one- or two-particle friction
coefficients can be used to obtain the bulk rheology.

\begin{acknowledgments}
We thank Bob Jones for pointing out an error in the first draft of
this paper, Peter Olmsted for useful discussions and Dr S.W.
Provencher for providing the CONTIN code. We gratefully
acknowledge financial support from the UK Engineering and Physical
Sciences Research Council.
\end{acknowledgments}


\end{document}